%% file: main.tex
\documentclass[10pt,twocolumn,letterpaper]{article}

\usepackage{cvpr}
\usepackage{amsmath}
\usepackage{amssymb}
\usepackage{graphicx}
\usepackage{bm}
\usepackage{booktabs}
\usepackage[table]{xcolor}
\usepackage{caption}
\usepackage{bbding}

\newcommand\blfootnote[1]{%
  \begingroup
  \renewcommand\thefootnote{}\footnote{#1}%
  \addtocounter{footnote}{-1}%
  \endgroup
}
\newcommand{\cmark}{\color{black}{\CheckmarkBold}}
\newcommand{\xmark}{\color{black}{\XSolidBrush}}

\definecolor{cvprblue}{rgb}{0.21,0.49,0.74}
\usepackage[pagebackref,breaklinks,colorlinks,allcolors=cvprblue]{hyperref}

\def\paperID{42831}
\def\confName{CVPR}
\def\confYear{2026}

\title{Scene-Level Heterogeneous Physics Simulation with 3D Gaussian Splats}

\author{
 Xiaoyang Liu$^{1}$ \qquad Shangzhe Wu$^{2, \dagger}$ \qquad Kai Han$^{1, \dagger}$ \\
 $^1$The University of Hong Kong \qquad $^2$University of Cambridge \\
 {\tt\small xiaoyangliu@connect.hku.hk \qquad sw2181@cam.ac.uk \qquad kaihanx@hku.hk} \\
 \href{https://visual-ai.github.io/raf/}{https://visual-ai.github.io/raf/}
}

\begin{document}

\twocolumn[{%
\renewcommand\twocolumn[1][]{#1}%
\maketitle

\begin{center}
    \vspace{-2em}
    \captionsetup{type=figure}
  \includegraphics[width=\textwidth]{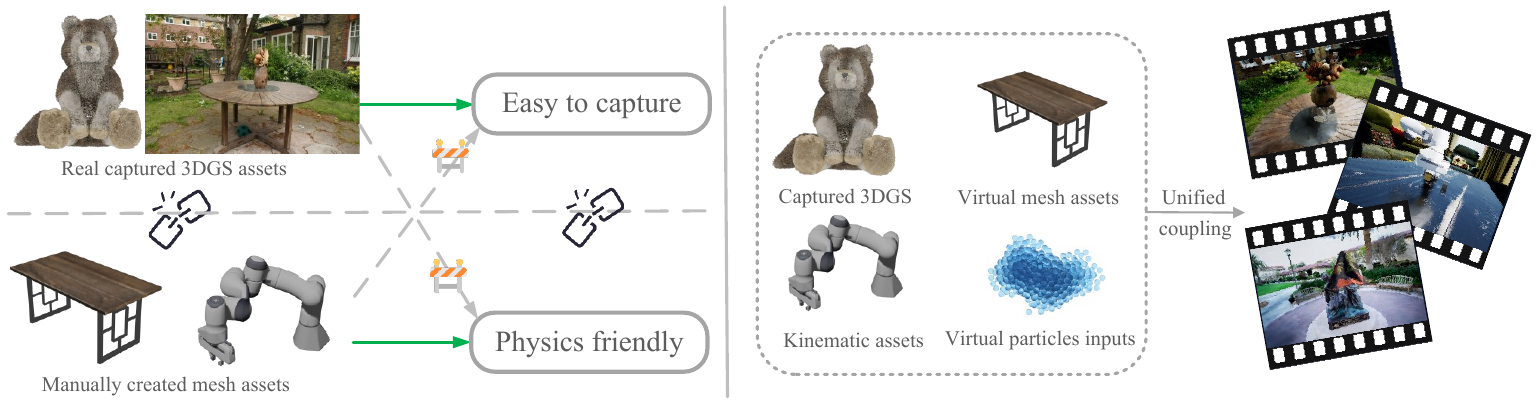}
  \captionof{figure}{
  \textbf{Bridging the representation gap.} \textit{Left}: Photorealistic 3DGS assets are easy to capture but incompatible with physics engines, while traditional meshes are physics-ready but lack capture fidelity. \textit{Right}: Our representation abstraction framework unifies heterogeneous assets — captured 3DGS, virtual meshes, and particle systems — within a single scene-level physical simulation.
}
  \label{fig:teaser}
\end{center}
}]
\blfootnote{$^{\dagger}$Corresponding authors.}

\input{sec/abstract}
\input{sec/introduction}
\input{sec/related_work}
\input{sec/method}
\input{sec/experiments}
\input{sec/conclusion}

{
    \small
    \bibliographystyle{ieeenat_fullname}
    \bibliography{main}
}

\setcounter{section}{0}
\setcounter{figure}{0}
\setcounter{table}{0}
\setcounter{equation}{0}

\input{sec/supplementary}

\end{document}

%% file: sec/abstract.tex
\begin{abstract}
3D Gaussian Splatting (3DGS) has achieved state-of-the-art photorealistic rendering, but the representation gap prevents these assets from being physically interactive. Production-grade physics engines do not understand the 3DGS representation, while prior physics-for-3DGS methods are monolithic silos. These prior works are fundamentally limited, demonstrating only object-centric physics in isolated environments, such as on an ideal plane — they are incapable of interacting with complex static collision geometry or heterogeneous assets. We propose a novel framework that, for the first time, bridges this gap by enabling 3DGS assets to participate in scene-level, heterogeneous, multi-solver physical simulations. Our core contribution is a Representation Abstraction Framework that ``translates" all diverse assets---including 3DGS, virtual meshes, and fluids---into a unified physical particle set. This abstraction is key to enabling complex behaviors, such as the non-rigid deformation of 3DGS assets, within a unified physics pipeline. This particle set, along with the static scene collision boundaries derived from scene capture, is processed within a solver-agnostic physics kernel. The physical results are then mapped back to drive each asset's specific visual reconstruction. This architecture unlocks capabilities impossible with prior art. We demonstrate complex, two-way interactions between deformable 3DGS assets, standard CG assets (like fluids and meshes), and large-scale captured static environments, showcasing realistic coupled phenomena that were previously unattainable.
\end{abstract}

%% file: sec/introduction.tex
\section{Introduction}
\label{sec:intro}

\begin{table*}[t]
\centering
\caption{\textbf{Comparison of physics-based 3D Gaussian splatting methods.} Our framework unifies high-fidelity captured objects (GS particles) with external standard assets (meshes/fluids) in a single scene-level simulation with advanced rendering.
}
\label{tab:comparison}
\resizebox{\textwidth}{!}{%
\begin{tabular}{@{}lccccc@{}}
\toprule

& PhysGaussian \cite{physgaussian} & DecoupledGaussian \cite{degs}  & Gaussian Splashing \cite{gaussiansplashing} & GausSim \cite{gaussim} & \textbf{Ours} \\
\midrule
Core paradigm & Unified (MPM) & Decoupled \& inpaint & Unified (PBD) & Unified (neural) & Hybrid representation \\
Dynamic object rep. & GS particles only & GS particles only & GS particles only & GS particles only & Hybrid (GS + mesh/fluid) \\
\midrule
Sim. captured objects & \cmark & \cmark & \cmark & \cmark & \cmark \\
Sim. external assets & \xmark & \xmark & \xmark & \xmark & \cmark \\
Heterogeneous interaction & \xmark & \xmark & \xmark & \xmark & \cmark (e.g., GS $\leftrightarrow$ mesh) \\
\midrule
Physics solver generality & Single (MPM) & Single (MPM) & Single (PBD) & Single (neural) & Multiple (MPM, SPH, rigid...) \\
Scene-level interaction & \xmark & \cmark & \xmark & \xmark & \cmark \\
Complex terrain support & \xmark (Ideal plane) & \xmark (Ideal plane) & \xmark (Ideal plane) & \xmark (Ideal plane/void) & \cmark (Full 3D) \\
\midrule
Rendering pipeline & Custom rasterizer & Custom rasterizer & Custom (PBR) & Custom rasterizer & Industrial (Unreal Engine 5) \\
Advanced lighting/shadows & \xmark & \xmark & Limited & \xmark & \cmark (Lumen, RT) \\
\bottomrule
\end{tabular}%
}
\end{table*}

3D Gaussian splatting (3DGS) \cite{3dgs} has recently become the state-of-the-art (SOTA) for real-time, photorealistic novel-view synthesis. It represents scenes with millions of explicit Gaussian splats, achieving unprecedented rendering quality and speed. However, while 3DGS has seen immense success in rendering static scenes or replaying pre-captured motion, these high-fidelity worlds remain largely ``dead'' \cite{vrgs,robotics}---they lack physical interactivity. Transforming these assets from ``passive renderables" to ``active participants" is the next major frontier to unlock their full potential \cite{survey2}.

The core challenge in achieving this stems from a fundamental representation gap \cite{physgaussian}. On one hand, 3DGS is a powerful visual representation defined by covariance ($\Sigma$), spherical harmonics ($\mathcal{C}$), and opacity ($\sigma$). On the other hand, powerful, mature, production-grade physics engines (e.g., Unreal Engine \cite{ue}, Houdini \cite{houdini}) are representation-agnostic, operating on a completely different set of physical representations (like particles, meshes, and rigid bodies) and supporting heterogeneous, multi-solver (MPM \cite{mpm}, SPH \cite{sph}, PBD \cite{pbd}) architectures. These engines cannot natively understand the novel 3DGS representation.

To bridge this chasm, prior work was forced down a ``siloed'' path \cite{physgaussian, gasp, physdreamer, physics3d, physsplat}: they built proprietary, monolithic physics solvers specifically for 3DGS. This approach, however, suffers from two fundamental flaws. First, it is homogeneous: it creates a non-extensible closed system in which only GS particles exist, making it fundamentally incapable of interacting with external, standard CG assets (like meshes, fluids, or rigid bodies). Second, it is isolated: it is not scene-level. It typically operates in a void or on an ideal plane and is fundamentally unable to support complex, pre-existing 3D static collision geometry. This ``toy'' approach is a dead end for scalable and general physics in complex worlds \cite{survey3, survey}.

In this paper, we argue against re-inventing the wheel with another siloed solver. Instead, we propose a novel \emph{Representation Abstraction Framework}. Our framework acts as a bridge that, for the first time, unifies 3DGS captured assets, traditional CG assets, and complex static scenes within powerful, ``off-the-shelf" heterogeneous physics engines.

This framework aims to solve both flaws of the siloed path. First, to solve the homogeneous problem, we translate and abstract all diverse dynamic assets---whether 3DGS (as soft bodies), virtual meshes (as PBD deformable bodies), or external fluids (as SPH particles)---into a unified physical particle set. Second, to solve the isolated problem, we also consume the static scene boundaries ($M_{\text{static}}$), which are extracted from the scene capture \cite{degs}.

These two elements---the unified dynamic particles and the static scene boundaries---are then fed into a unified simulation context. Here, a modular heterogeneous solver handles all physical interactions between assets~\cite{newton} and the static scene (e.g., particle-mesh collision, fluid-on-soft-body coupling). After simulation, our framework ``reverse-translates'' the physical results (e.g., deformation gradient $F'$, new position $x'$) and intelligently routes them to each asset's own specific visual reconstruction pipeline: 3DGS assets use $F'$ to update their covariance $\Sigma'$, while mesh assets use particle positions to drive barycentric skinning.

We demonstrate this end-to-end system by validating it on a series of previously impossible, complex dynamic scenarios. For example, we show a captured 3DGS soft-body bear deforming under a virtual SPH fluid splash while realistically sliding across a captured static garden scene with complex geometry. Our work ultimately transforms 3DGS from static renderables into fully functional, interactable ``physical sandboxes'', achieving dual physical and visual high-fidelity within a unified production renderer like Unreal Engine 5 \cite{ue}.

In this paper, we make the following contributions: (\textit{i}) We propose the first framework to enable scene-level, heterogeneous, multi-solver physics for 3D Gaussian Splatting. (\textit{ii}) We introduce a representation abstraction layer that translates diverse dynamic assets (3DGS, meshes, fluids) and complex static scenes into a single physical simulation context. (\textit{iii}) We demonstrate complex, two-way physical interactions between captured 3DGS assets, standard CG assets, and real-world static scene geometry. (\textit{iv}) We deliver a complete end-to-end pipeline with unified rendering in Unreal Engine 5, leveraging advanced features such as Lumen global illumination.

%% file: sec/related_work.tex
\section{Related work}
\label{sec:related}

\subsection{Physics simulation of 3D Gaussian splatting}
\label{sec:related_gs_physics}

Despite 3DGS's rapid adoption for high-fidelity rendering, prior efforts~\cite{physgaussian, dreamphysics, gasp, gaussiansplashing} to endow it with physics have followed a \textit{monolithic and siloed} paradigm.
These works fall into two main categories: analytical solvers (e.g., PhysGaussian \cite{physgaussian} and others treating kernels as material points in MPM or adapting them for PBD) and neural solvers (e.g., GausSim \cite{gaussim}, which uses a center-of-mass systems model \cite{springmass} governed by continuum mechanics to learn dynamics).

While groundbreaking, this entire monolithic approach suffers from fundamental limitations that restrict its use to ``toy'' scenarios \cite{survey2, degs}. First, it locks the asset into a single-solver model, preventing complex multi-material simulation. Second, and more critically, these GS-specific solvers are siloed and non-extensible. They are incapable of interacting with external, standard virtual assets (such as meshes, fluids, or rigid bodies). Consequently, these methods are demonstrated in isolated environments, such as on an ideal plane, and cannot support complex, pre-existing 3D static collision geometry. In summary, the field has been missing a general and scalable solution to integrate high-fidelity GS assets into complex, scene-level simulations. Our work directly confronts this limitation.

\subsection{Heterogeneous multi-solver physics}
\label{sec:related_heterogeneous_physics}

In the broader computer graphics community, high-fidelity physical simulation for production has long embraced a heterogeneous, multi-solver architecture \cite{houdini, ue,genesis,newton}. The power of modern physics engines lies not in a single, ``one-size-fits-all'' solver, but in the ability to couple multiple, specialized solvers, managing complex interactions between assets represented in entirely different ways (e.g., coupling SPH for fluids, MPM for granulars, and PBD for cloth). The core capability of these industrial-strength systems is a unified collision and coupling mechanism that manages interactions between these disparate solvers as a generalized process.

These powerful engines are, however, representation-agnostic. They operate on a common currency of particles and explicit triangle meshes \cite{gasp}. They have no native understanding of rendering-oriented representations such as  NeRF \cite{nerf} or 3DGS \cite{3dgs,2dgs}. Therefore, a significant gap exists. Instead of ``re-inventing the wheel'' by building another limited, GS-specific solver, the more intelligent and scalable path is to bridge this gap. Our core contribution, a \emph{Representation Abstraction Framework}, is precisely this bridge.

\subsection{Neural scene representations}
\label{sec:related_representations}

While implicit representations like NeRF \cite{nerf} first achieved photorealism, their monolithic structure and slow rendering make them extremely difficult to physically simulate. This limitation led to the shift towards explicit or hybrid representations, culminating in 3D Gaussian splatting, whose explicit, point-based nature is far more suitable for physical interaction than an implicit field. However, recent works on dynamic 3DGS (or 4DGS \cite{4dgs,dreamgaussian4d}) focus predominantly on kinematic motion. These methods (e.g., \cite{gaussianflow}) only replay a pre-captured or generated motion sequence. They are not physics-based; they cannot simulate new dynamics or respond to novel physical interactions. Simulating new, physically plausible dynamics for these assets remains a critical open challenge, which our work is the first to tackle within a general-purpose, heterogeneous, and scene-level simulation framework.

%% file: sec/method.tex
\section{Method}
\label{sec:methodology}

\begin{figure*}[t]
    \centering
    \includegraphics[width=\textwidth]{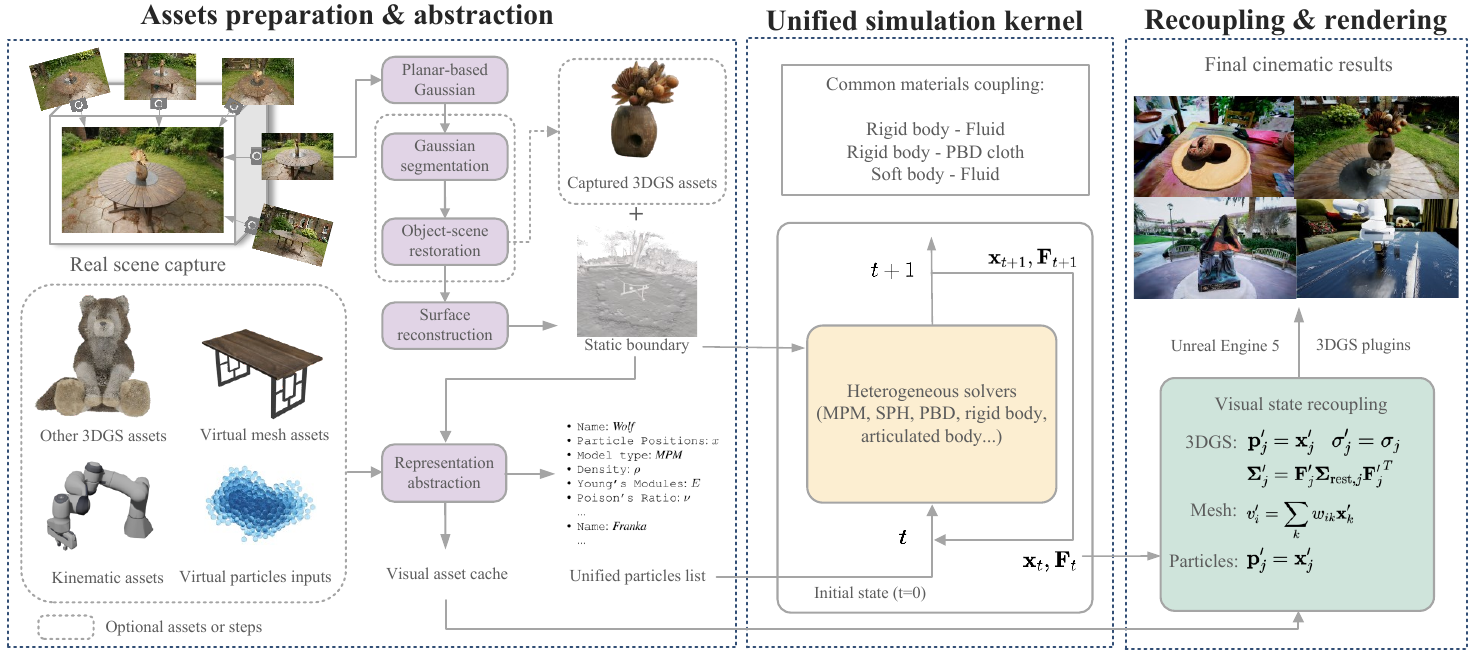}
    \caption{
        \textbf{Architecture of our Representation Abstraction Framework}.
        \textit{(1) Asset preparation \& abstraction.}
        Heterogeneous inputs (left), including real scene capture and various imported assets, are processed. A static collision mesh ($M_{\text{static}}$) is reconstructed from multi-view images using planar-based Gaussian. The representation abstraction layer translates all assets into a unified particle list. Visual-only data (e.g., \{$\Sigma_{\text{rest}}$, $w_{ik}$\}) is cached.
        \textit{(2) Unified simulation kernel.}
        The kernel consumes the particles and static mesh, using a suite of heterogeneous solvers (MPM, SPH, PBD-cloth, rigid-body, articulated-body) to output the updated physical state ($\mathbf{x}'$, $\mathbf{F}'$).
        \textit{(3) Heterogeneous recoupling \& rendering.}
        The visual state recoupling step ``reverse-translates'' the physical state using the cache. It performs asset-specific reconstruction and serialization. These are loaded into Unreal Engine 5 to produce the final cinematic results.
    }
    \label{fig:pipeline}
\end{figure*}

We introduce a novel end-to-end framework that enables captured 3D Gaussian splatting (GS) assets to participate in complex, scene-level physical simulations with heterogeneous virtual assets. The core challenge is that modern physics engines are not designed to process GS representations, while existing GS-specific physics methods are siloed and cannot interact with complex, multi-solver scenes.

Our key contribution is a \textit{Representation Abstraction Framework}, which acts as a bridge to resolve this. As shown in Figure \ref{fig:pipeline}, our framework is organized into three stages: (1) heterogeneous asset preparation and abstraction, (2) a unified simulation kernel, and (3) heterogeneous visual recoupling and rendering.

\subsection{Asset preparation and abstraction}
\label{sec:stage1}

The goal of this stage is to consume a diverse set of optional input assets (shown in Fig. \ref{fig:pipeline}, left) and translate them into the two key components required by our simulation kernel: a single static collision mesh ($\mathcal{M}_{\text{static}}$) and a single unified particle list.

\paragraph{Populating the static world ($\mathcal{M}_{\text{static}}$).}
The $\mathcal{M}_{\text{static}}$ defines all non-moving geometry our dynamic objects can interact with. It is a composite mesh assembled from all static sources, including both captured real-world data and optional imported assets. For in-the-wild scenes, we process multi-view images \cite{mip, mip2, tnt} through our captured asset pipeline. This pipeline uses techniques like Gaussian segmentation \cite{seganygs} and object-scene restoration (leveraging planar-based Gaussian \cite{pgsr} and surface reconstruction \cite{poisson}) to extract the base static geometry from the real scene capture. Users can also augment the static world with their own custom geometry, such as imported 3D Gaussian splatting (3DGS) assets or virtual mesh assets. All selected static sources are processed and merged into the final watertight $\mathcal{M}_{\text{static}}$.

\noindent \textbf{Populating the dynamic world (unified particle list).}
All dynamic participants are processed by our representation abstraction layer and converted into a single unified particle list. This layer handles a wide variety of asset types:

\textit{Captured 3DGS assets:} Dynamic GS objects ($\mathcal{G}_{\text{dyn}}$) are converted into a set of physical particles $\mathcal{X}_{\text{gs}}$.
If the asset has been pre-processed by the object-scene restoration pipeline of DecoupledGaussian~\cite{degs},  we use its dense points directly. Otherwise, for standard 3DGS captures, we define the object's interior by computing a continuous opacity field $d(\mathbf{x})$ \cite{physgaussian} and sampling it:
\begin{equation}
d(\mathbf{x}) = \sum_{g \in \mathcal{G}_{\text{asset}}} \sigma_{g} \exp\left(-\frac{1}{2}(\mathbf{x}-\mathbf{k}_{g})^{T}\mathbf{\Sigma}_{g}^{-1}(\mathbf{x}-\mathbf{k}_{g})\right)
\label{eq:opacity}
\end{equation}
For each resulting particle position $\mathbf{x}_j$, we instantiate a physical entity $p_j$. This entity, which is consumed by the simulation kernel, encapsulates all necessary information: its geometric and kinematic state (e.g., position $\mathbf{x}_j$, mass $m_j$, velocity $\mathbf{v}_j$), its assigned physical model ($\texttt{ModelType}_j$), and its constitutive parameters (e.g., Young's modulus $E$, Poisson's ratio $\nu$ for MPM particles). The full structure is defined as:
\begin{equation}
p_j = \{ \mathbf{x}_j, m_j, \mathbf{v}_j, \texttt{ModelType}_j, E_j, \nu_j, \dots \}
\label{eq:physical_entity}
\end{equation}
Concurrently, the visual entity $v_j = \{ \mathbf{\Sigma}_{\text{rest}, j}, \mathcal{C}_j, \sigma_{j} \}$ is stored in the visual asset cache for later recoupling.

\textit{Virtual meshes:} Dynamic meshes are ``particle-ized'' using Poisson disk sampling.
\begin{equation}
\forall \mathbf{x}_i, \mathbf{x}_k \in \mathcal{X}_{\text{mesh}}, \quad ||\mathbf{x}_i - \mathbf{x}_k|| \ge r \quad \text{for } i \neq k
\label{eq:poisson}
\end{equation}
We establish a binding \cite{skinning} between the template mesh $\mathcal{M}_{\text{template}}$ and its particles by pre-computing barycentric weights $\{w_{ik}\}$, such that its rest position is defined by:
\begin{equation}
v_i^{\text{rest}} = \sum_{k} w_{ik} \mathbf{x}_k
\label{eq:barycentric}
\end{equation}
This template mesh and its weights are also stored in the visual asset cache.

\textit{External particle systems (e.g., fluid):} These inputs (Virtual Particles) are loaded directly as a set of initial particle positions $\mathcal{X}_{\text{particle}}$. Their $\texttt{ModelType}$ is set (e.g., to SPH), and no separate visual entity is required.

\textit{Kinematic assets (e.g., robot arm \cite{franka}):} These assets are not converted to particles. They are treated as standard virtual mesh assets (rigid bodies) whose motion is not solved by our physics kernel. Instead, their transforms are kinematically prescribed by the user-defined animation. They act as one-way colliders, affecting other dynamic objects (like the rigid-body cubes) while remaining unaffected by them.

\subsection{Unified simulation kernel}
\label{sec:stage2}

This stage acts as a modular, solver-agnostic physics kernel. As shown in Figure \ref{fig:pipeline}, the kernel takes two unified inputs from Stage 1: the complete static collision mesh ($\mathcal{M}_{\text{static}}$) and the unified particle list at the current time $t$.

The particles are initialized with their current physical state ($\mathbf{x}_t, \mathbf{F}_t$), derived from the initial state ($t=0$) or the previous loop's output. These inputs are processed by the heterogeneous solvers module, which manages and seamlessly interfaces with various specialized solvers (e.g., MPM, SPH, PBD-cloth, rigid body, and articulated body).

The kernel's core capability is its ability to robustly manage heterogeneous interactions. This includes handling complex, continuous force exchanges (e.g., pressure from SPH particles onto MPM particles) as well as discrete collision events.

For discrete collision and frictional contact between all entities (e.g., particle-mesh or rigid-body-on-rigid-body), the kernel employs a unified impulse-based mechanism. As an example of this discrete coupling mechanism, when two entities $i$ and $j$ collide, a response impulse is computed to determine the post-collision velocities.
Let $r_n$ and $\mathbf{r}_t$ denote the normal and tangential components of the relative velocity between the two colliding entities. The post-collision velocity components are determined by:
\begin{align}
r_n^* &= -e r_n \label{eq:normal_response} \\
\mathbf{r}_t^* &= \mathbf{r}_t \max(0, 1 - \mu |r_n| / |\mathbf{r}_t|) \label{eq:tangent_response}
\end{align}
where $e$ is the coefficient of restitution and $\mu$ is the Coulomb friction coefficient.

This unified impulse-based approach ensures non-penetrating and frictional response across all dynamic assets when they collide with each other or with $\mathcal{M}_{\text{static}}$. Within each time step, the heterogeneous solvers are executed iteratively in a specific sequence — rigid-body and articulated-body solvers first, followed by MPM, SPH, and finally PBD —
with updated particle states propagated between solvers and refined through inter-solver iterations, ensuring that complex coupling forces and constraints are consistently resolved within the same step.
More detailed solver and continuous coupling implementations are provided in Supplementary S6.
At the end of the step $\Delta t$, the simulation kernel returns the updated physical state ($\mathbf{x}_{t+1}, \mathbf{F}_{t+1}$), which is passed to the final stage for visual reconstruction.

\subsection{Heterogeneous recoupling and rendering}
\label{sec:stage3}

This final stage reverse-translates the sparse, abstract physical data from the simulation kernel back into high-fidelity, renderable visual assets. This stage requires two inputs: the updated physical state ($\mathbf{x}', \mathbf{F}'$) from Stage 2, and the visual asset cache (containing $\{ \mathbf{\Sigma}_{\text{rest}}, w_{ik}, ... \}$) from Stage 1.

\paragraph{Visual state recoupling.}
The visual state recoupling step performs asset-specific reconstruction in parallel for all dynamic entities. As shown in Figure \ref{fig:pipeline}, different asset types use different reconstruction logic:

\begin{itemize}
    \setlength\itemsep{0em}
    \item \textit{3DGS assets:} We propagate the physical deformation to the visual attributes.
    The new Gaussian center $\mathbf{p}'_j$ is set to the updated particle position.
    \begin{equation}
        \mathbf{p}'_j = \mathbf{x}'_j
    \label{eq:pos_update}
    \end{equation}
    The new covariance $\mathbf{\Sigma}'_j$ is computed by applying the physical deformation $\mathbf{F}'_j$ to the cached rest-state covariance $\mathbf{\Sigma}_{\text{rest},j}$:
    \begin{equation}
        \mathbf{\Sigma}'_j = \mathbf{F}'_j \mathbf{\Sigma}_{\text{rest}, j} {\mathbf{F}'_j}^T
    \label{eq:cov_update}
    \end{equation}
    The opacity $\sigma'_{j}$ is assumed to be an invariant material property and is carried over from its rest state:
    \begin{equation}
        \sigma'_{j} = \sigma_{j}
    \label{eq:opacity_update}
    \end{equation}
    Finally, the rotation (extracted from $\mathbf{F}'_j$) is used to rotate the SH coefficients $\mathcal{C}_j$. The mathematical derivation proving this $F$ propagation to $\Sigma$ is physically consistent is provided in Supplementary S3.

    \item \textit{Mesh assets:} We retrieve the asset's template mesh $\mathcal{M}_{\text{template}}$ and its precomputed barycentric weights $\{w_{ik}\}$ from the cache. The new deformed vertex position $v'_i$ is computed by applying these same weights to the new particle positions $\mathbf{x}'_k$:
    \begin{equation}
        v'_i = \sum_{k} w_{ik} \mathbf{x}'_k
    \label{eq:skinning_update}
    \end{equation}

    \item \textit{Particle assets:} For fluid assets (SPH), the physical particle position is its own visual representation: $\mathbf{p}'_j = \mathbf{x}'_j$.
\end{itemize}

\paragraph{Serialization and final rendering.}
The recoupled visual data is serialized into asset-aware formats (e.g., \texttt{.ply} files for 3DGS, \texttt{.abc} files for meshes, \texttt{.pbd} caches for particles). These serialized files are then loaded into our final cinematic results pipeline. We utilize Unreal Engine 5 as our final renderer. The engine natively ingests the \texttt{.abc} and \texttt{.pbd} caches, while our 3DGS plugins \cite{volinga} (compatible Gaussian Splatting rendering components) load the per-frame \texttt{.splat} files. This final step allows all our co-simulated assets to be rendered together, interacting correctly with advanced features like Lumen global illumination and ray-traced shadows.

%% file: sec/experiments.tex
\section{Experiments}
\label{sec:experiments}

\begin{figure*}[t]
    \centering
    \includegraphics[width=\textwidth]{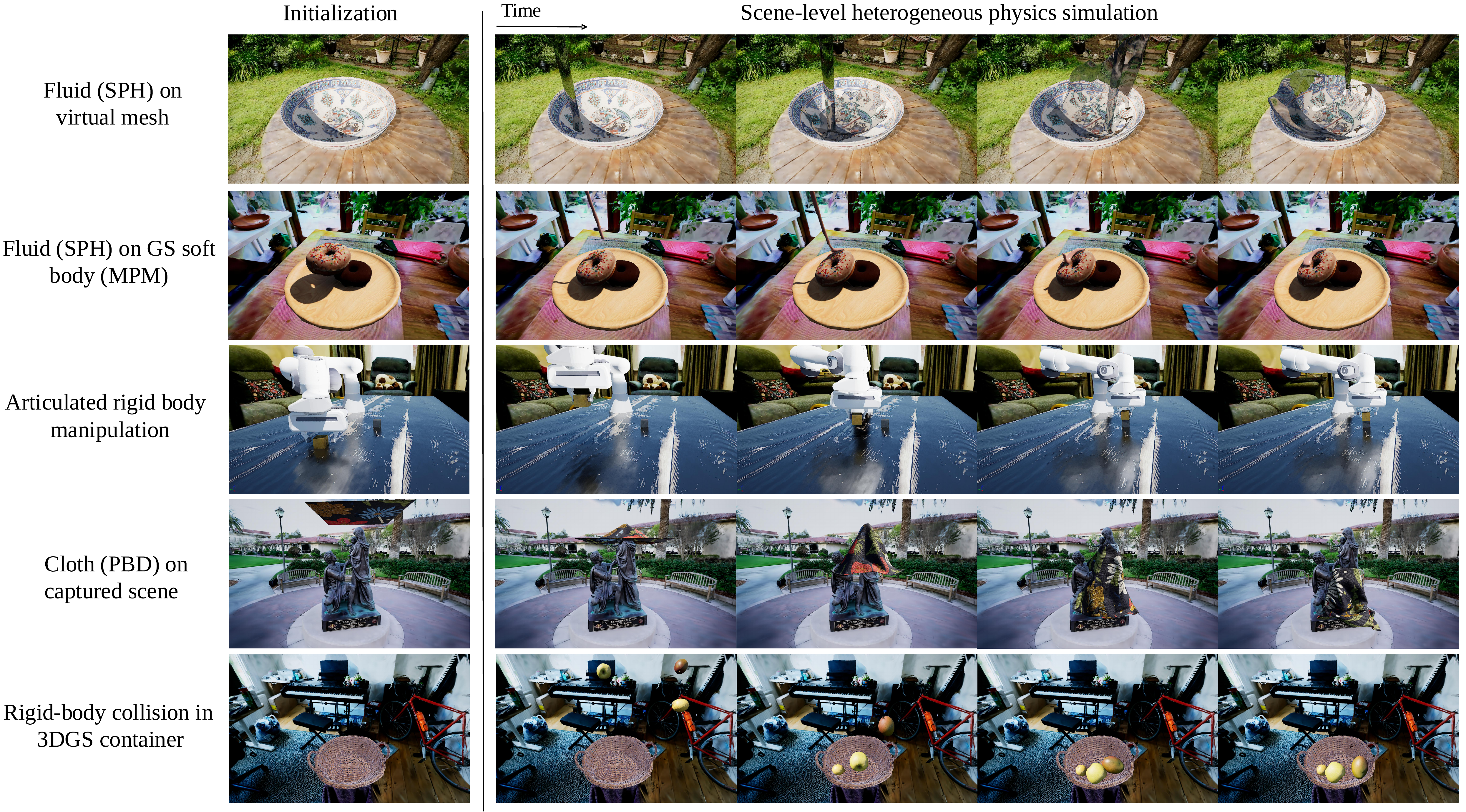}
    \caption{
        \textbf{Our qualitative showcase of new, scene-level heterogeneous simulations produced using our framework.} Each row demonstrates a unique coupling of different asset types and physics solvers, executed within a captured 3DGS environment and rendered in Unreal Engine 5.
        \textit{Row 1: Fluid (SPH) on virtual mesh.} A virtual fluid (SPH) is poured into an imported virtual mesh (bowl), interacting with both the bowl and the static 3DGS scene (garden table).
        \textit{Row 2: Fluid (SPH) on GS soft body (MPM).} A high-viscosity fluid (SPH) is coupled with a captured 3DGS soft body (MPM donut), demonstrating complex multi-solver, two-way interaction on a plate.
        \textit{Row 3: Articulated rigid body manipulation.} A kinematically-driven robotic arm (articulated-body solver) manipulates a virtual rigid body (cube) on a table.
        \textit{Row 4: Cloth (PBD) on captured statue.} A virtual cloth (PBD solver) realistically drapes and slides over the complex, non-convex geometry of a captured 3DGS statue, refuting the ``ideal plane'' limitation.
        \textit{Row 5: Rigid-body collision in 3DGS container.} Multiple virtual rigid bodies (fruits) interact with each other and with the complex geometry of an imported 3DGS asset (the basket).
        These scenarios are, by design, unattainable by prior ``siloed'' methods.
    }
    \label{fig:qualitative_results}
\end{figure*}

We validate our framework across five complex scenarios that are, by design, unattainable by prior siloed methods. Each scenario exercises a distinct capability: scene-level composition, heterogeneous multi-solver coupling, articulated dynamics, complex geometry interaction, and integration with imported 3DGS assets — all while maintaining photorealistic rendering via Unreal Engine 5.

\subsection{Implementation details}

\paragraph{Simulation kernel.}
We implement our core abstraction layer and simulation loop (described in Sec. \ref{sec:methodology}). Our unified kernel, which is built upon the Genesis-world physics solver \cite{genesis}, acts as a modular, solver-agnostic kernel that manages and dispatches tasks to specialized solvers, including MPM (for soft bodies), SPH (for fluids), PBD (for cloth), standard rigid-body dynamics (for objects), and articulated-body dynamics (for kinematic chains).

\paragraph{Asset preparation.}
The preparation of static and dynamic assets follows the structure detailed in Section \ref{sec:stage1}. For scene-level static geometry, two strategies are employed to construct the static collision mesh ($\mathcal{M}_{\text{static}}$): direct capture and composition. For scenes captured in situ (e.g., \textit{Kitchen}, \textit{Statue}), we utilize our full asset pipeline to process multi-view images and extract the base static geometry ($\mathcal{M}_{\text{static}}$) from the real scene capture. For compositional scenarios (e.g., \textit{Garden}, \textit{Rooms}), we load a primary 3DGS scene and then import additional assets—such as virtual mesh bowls or pre-captured 3DGS fruit baskets—which are also processed and merged into the final static collision geometry. Dynamic assets are uniformly converted into the unified particle list, regardless of their original source format.

\paragraph{Rendering.}
All scenarios are serialized (e.g., \texttt{.splat}, \texttt{.abc} caches) and rendered in Unreal Engine 5 using a compatible 3DGS rendering component and leveraging Lumen for global illumination and shadows.

\paragraph{Baselines for comparison.}
Our primary baselines are the ``siloed'' SOTA methods (e.g., PhysGaussian \cite{physgaussian}, GausSim \cite{gaussim}).
Crucially, these baselines cannot be applied to our experimental settings.
They are not designed for heterogeneous, scene-level composition. Our qualitative showcase (Sec. \ref{sec:qualitative}) therefore demonstrates capabilities that are, by design, impossible for them.

\subsection{Qualitative showcase of new capabilities}
\label{sec:qualitative}

We present five complex scenarios, visualized in Figure \ref{fig:qualitative_results} and our supplementary video, to demonstrate the breadth of new capabilities unlocked by our framework. These scenarios are chosen to validate our core claims by showcasing: (1) scene composition (SPH on virtual mesh), (2) multi-solver coupling (SPH on MPM 3DGS), (3) articulated dynamics (robot arm), (4) complex geometry interaction (PBD-cloth on statue), and (5) interaction with imported 3DGS assets (rigid-bodies in 3DGS container). As the figure's caption details, these scenarios---which unify SOTA physics and rendering---are unattainable by prior siloed methods.

\paragraph{Summary of qualitative findings.}
These five scenarios, taken together, validate our core claims. They demonstrate the unification of a wide range of solvers and an expansive set of heterogeneous assets (captured GS scenes, captured GS objects, imported GS assets, and virtual meshes/particles). Critically, Row 1 and Row 5 in Figure \ref{fig:qualitative_results} demonstrate that our framework is a powerful composition engine for 3DGS, a capability far beyond the ideal plane limitations of prior ``siloed'' methods like PhysGaussian or GausSim.

\subsection{Ablation study and comparative analysis}
\label{sec:comparisons}

To scientifically validate the necessity and effectiveness of our core contributions, we conduct a series of rigorous ablation studies and visual comparisons.
As no prior method supports heterogeneous, scene-level simulation,
a direct numerical comparison with prior ``siloed'' methods (like PhysGaussian \cite{physgaussian} or GausSim \cite{gaussim}) is not applicable.
Instead, we ablate our own components to demonstrate their necessity.

\subsubsection{Ablation study}
\label{sec:ablation}

We isolate three core components of our framework: the unified coupling kernel, the scene-level geometry, and the representation abstraction.

\begin{figure}[t]
    \centering
    \includegraphics[width=\columnwidth]{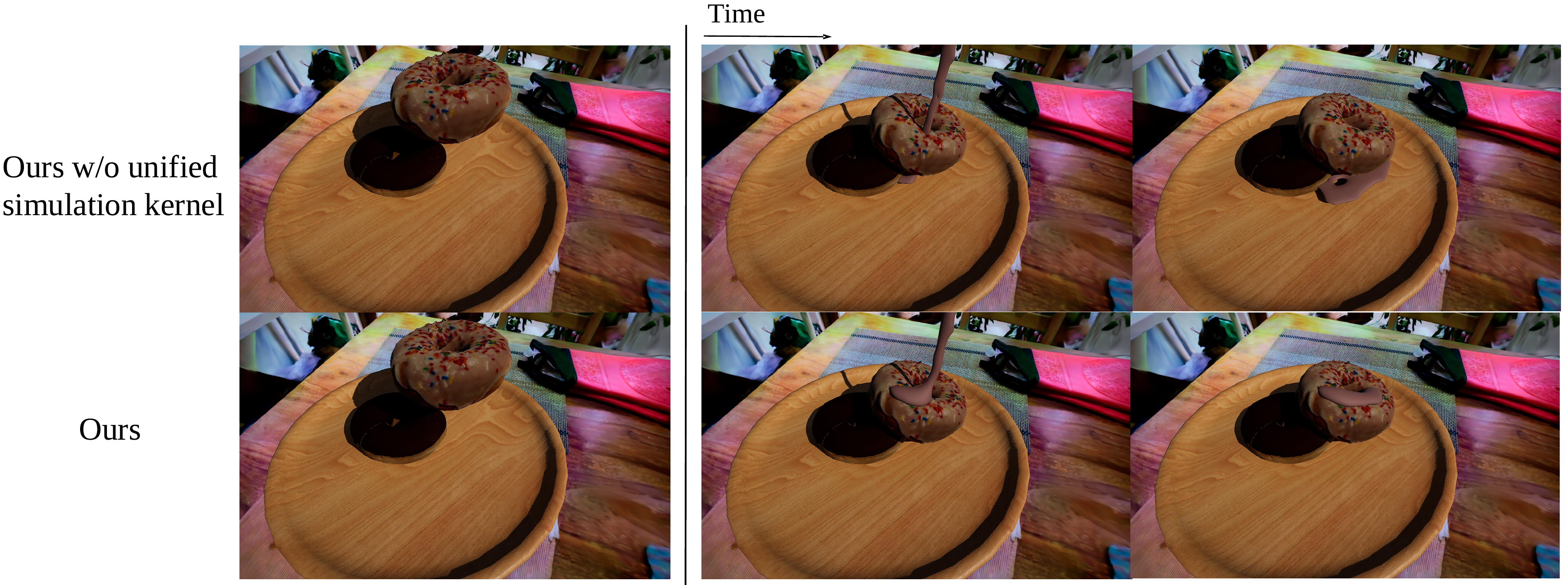}
    \caption{
        \textbf{Ablation study on our unified kernel.}
        \textit{Top row} (\texttt{Ours w/o unified simulation kernel}): Our ablated baseline. Without the heterogeneous coupling mechanism, the SPH fluid particles (e.g., \textit{Sauce}) and the MPM soft-body particles (e.g., \textit{Donut}) do not interact, passing through each other in a physically implausible manner.
        \textit{Bottom row} (\texttt{Ours}): Our full framework. The unified kernel correctly handles the SPH-on-MPM collision and coupling, allowing the sauce to realistically collide with, deform, and settle on top of the 3DGS \textit{Donut}.
    }
    \label{fig:ablation_kernel}
\end{figure}

\paragraph{Validating the unified kernel.}
To demonstrate the necessity of our heterogeneous coupling mechanism, we compare our full model against an ablated version, \texttt{Ours w/o unified simulation kernel}. In this baseline, the SPH and MPM solvers operate in the same space but do not interact. As visualized in Figure \ref{fig:ablation_kernel}, the ablated version (top row) fails, as the sauce and donut pass through each other, resulting in a physically implausible simulation. Our full model (bottom row) correctly handles the SPH-on-MPM coupling, where the (SPH) sauce collides with and deforms the (MPM) donut, demonstrating the critical importance of our unified kernel.

\begin{figure*}[htb]
    \centering
    \includegraphics[width=1\textwidth]{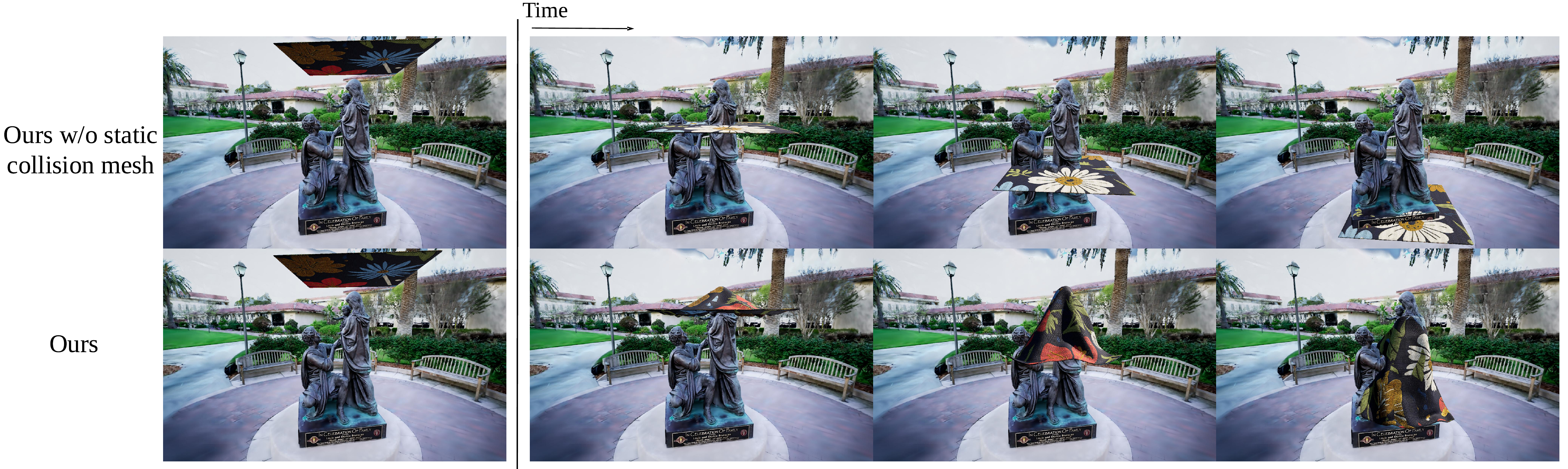}
    \caption{
        \textbf{Ablation study on scene-level geometry.}
        \textit{Top row} (\texttt{Ours w/o static collision mesh}): Our ablated baseline. Without the static collision mesh ($\mathcal{M}_{\text{static}}$), the cloth simulation is limited to an ``ideal plane'' and incorrectly passes through the statue.
        \textit{Bottom row} (\texttt{Ours}): Our full framework. The PBD-cloth solver correctly interacts with the complex, non-convex geometry of the captured 3DGS statue (represented by $\mathcal{M}_{\text{static}}$), resulting in realistic draping, folding, and sliding.
    }
    \label{fig:ablation_scene_level}
\end{figure*}

\paragraph{Validating scene-level interaction.}
To demonstrate that our framework has overcome the ``ideal plane'' limitation of prior ``siloed'' work, we compare our full model against an ablated baseline, \texttt{Ours w/o static collision mesh}. As visualized in Figure \ref{fig:ablation_scene_level}, the ablated baseline (top row) fails completely; without the complex $\mathcal{M}_{\text{static}}$ from our asset preparation workflow, the PBD-cloth solver defaults to an ``ideal plane'' and the cloth passes nonsensically through the captured statue. Our full method (bottom row) correctly uses the statue's high-fidelity collision mesh, resulting in realistic and complex draping behavior. This directly demonstrates that our scene-level pipeline is essential for integrating assets into captured environments.

\begin{figure}[t]
    \centering
    \includegraphics[width=\columnwidth]{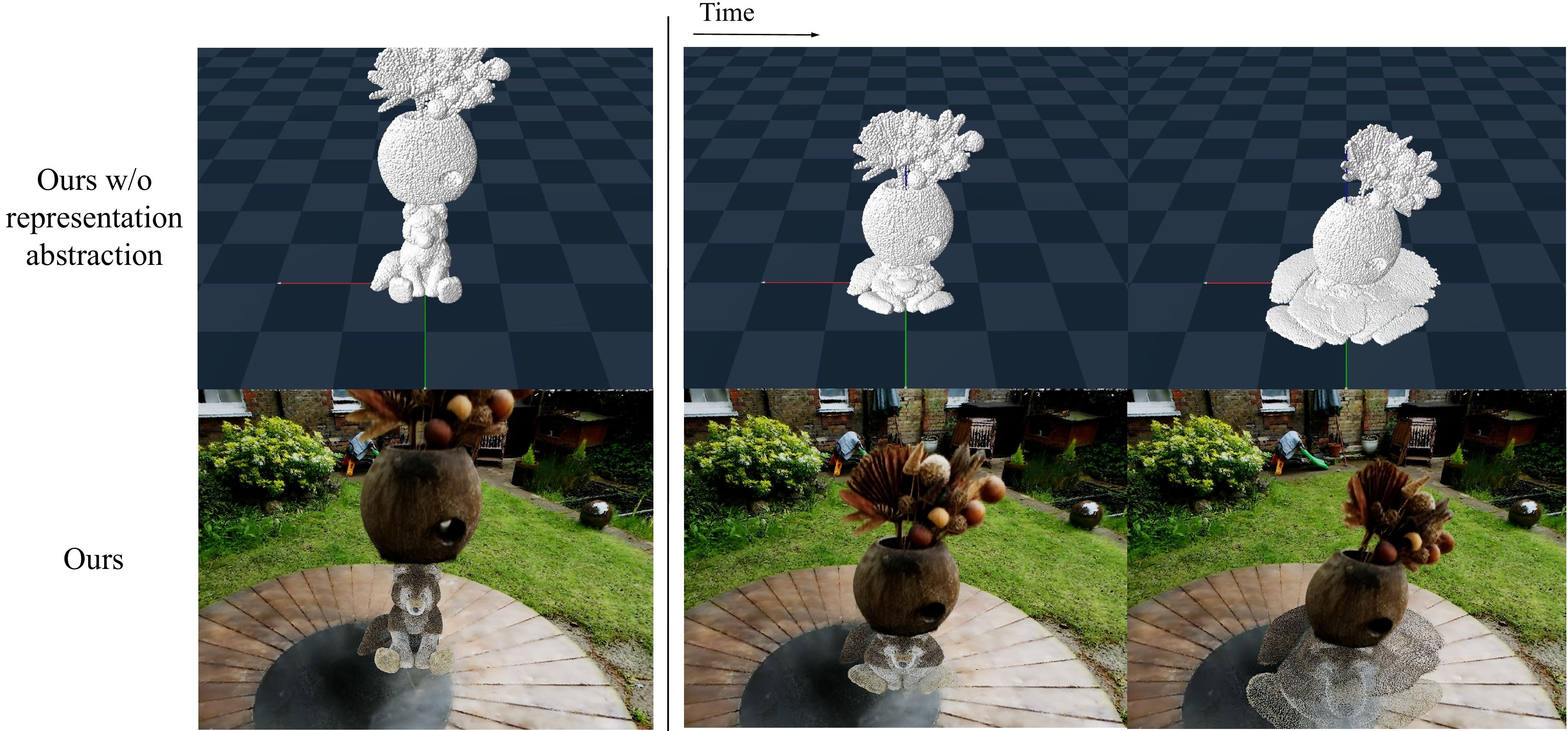}
    \caption{
\textbf{Ablation study on our representation abstraction.}
\textit{Top row} (\texttt{Ours w/o representation abstraction}): Without the representation abstraction, only the sparse simulation particles are visible, lacking continuous surfaces and photorealistic appearance. \textit{Bottom row} (\texttt{Ours}): Our full recoupling layer recovers high-fidelity visual attributes from the sparse physical state, demonstrating its necessity for photorealistic rendering.
    }
    \label{fig:ablation_abstraction}
\end{figure}

\paragraph{Validating representation abstraction.}
Finally, we validate the importance of our visual recoupling step. We compare our full result against \texttt{Ours w/o representation abstraction}. As shown in Figure \ref{fig:ablation_abstraction}, the top row visualizes only the sparse, underlying simulation particles, failing to convey the photorealistic appearance of the asset. Our full method demonstrates how our recoupling ``bridge" is essential for translating this sparse physical simulation back into a high-fidelity visual result.

\subsubsection{Comparison with a photogrammetry pipeline}
\label{sec:visual_comparison}

Having validated our pipeline's internal components, we now validate our core motivation for using 3DGS. One might ask: why not use standard photogrammetry, which also produces physics-ready meshes from images? We demonstrate the fundamental trade-off addressed by our framework.

\begin{figure}[t]
    \centering
    \includegraphics[width=\columnwidth]{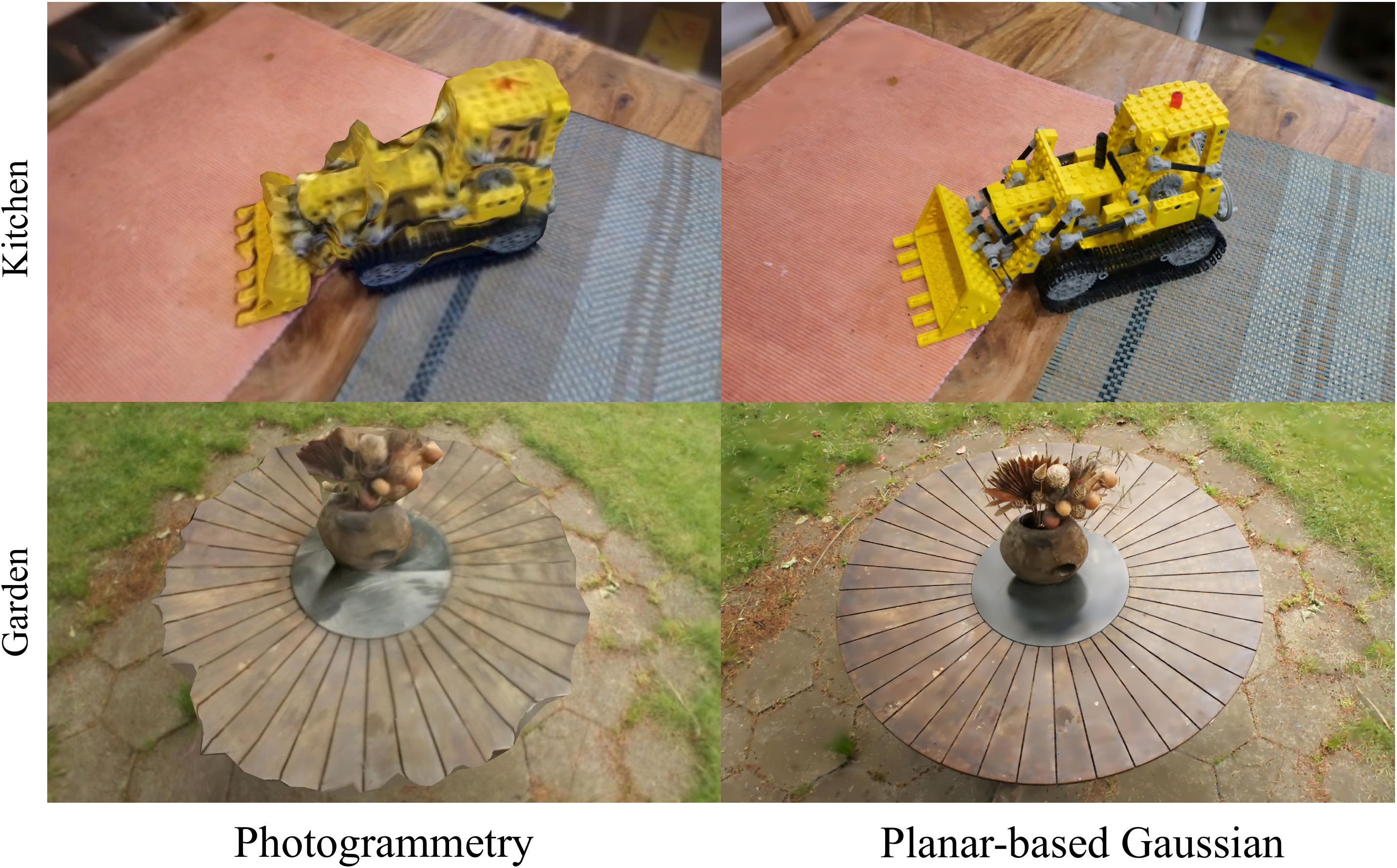}
    \caption{
        \textbf{Comparison with photogrammetry.}
        \textit{Left (Photogrammetry):} Standard mesh extraction from images suffers from severe geometric artifacts and missing details on fine structures.
        \textit{Right (Ours):} By leveraging 3DGS, our pipeline preserves sharp, complex geometries, which our RAF further transforms from passive renderables into interactive physical assets.
    }
    \label{fig:photogrammetry_comparison}
\end{figure}

We compare our method against a baseline generated from the same multi-view images using KiriEngine \cite{kiri}. As shown in Figure \ref{fig:photogrammetry_comparison}, the photogrammetry baseline (left) exhibits severe geometric artifacts, appearing ``blobby" or distorted on complex structures like bulldozer treads and table slats. In contrast, our pipeline (right), leveraging PGSR \cite{pgsr}, preserves sharp and geometrically complete details. This demonstrates that while traditional meshes are natively physics-ready, they lack the fidelity required for complex scenes. Our framework bridges this gap, enabling high-fidelity 3DGS assets to participate in advanced physics simulations for the first time.

%% file: sec/conclusion.tex
\section{Conclusion}
\label{sec:conclusion}
In this paper, we introduced the \textit{Representation Abstraction Framework} (RAF), the first end-to-end pipeline to bridge the fundamental gap between 3DGS assets and industrial-strength heterogeneous physics engines. Our core innovation is an abstraction layer that unifies diverse visual representations into a single physical particle set consumable by a multi-solver kernel (MPM, SPH, PBD, etc.). This architecture overcomes the limitations of prior ``siloed'' physics-for-GS methods, enabling scene-level composition, complex two-way heterogeneous coupling, and realistic interaction with complex static geometry — all rendered with photorealistic fidelity in Unreal Engine 5. Our ablation studies confirmed the necessity of the unified kernel, scene-level geometry handling, and visual recoupling bridge. Currently, our pipeline targets offline, cinematic-quality production, where simulation accuracy and visual fidelity are prioritized over runtime performance; extending it to real-time interactive scenarios and improving object-level separation in complex captured scenes are promising directions for future work.

\paragraph{Acknowledgement}
This work is supported by Hong Kong Research Grant Council - General Research Fund (Grant No. 17213825) and HKU Seed Fund for PI Research.

%% file: sec/supplementary.tex
\clearpage
\onecolumn

% Supplement numbering
\renewcommand{\thesection}{S\arabic{section}}
\renewcommand{\thesubsection}{\thesection.\arabic{subsection}}
\renewcommand{\thesubsubsection}{\thesubsection.\arabic{subsubsection}}

\renewcommand{\thefigure}{S\arabic{figure}}
\renewcommand{\thetable}{S\arabic{table}}
\renewcommand{\theequation}{S\arabic{equation}}

% IMPORTANT: unique hyperref anchors for supplement
\makeatletter
\@ifpackageloaded{hyperref}{
  \renewcommand{\theHsection}{supp.section.\arabic{section}}
  \renewcommand{\theHsubsection}{supp.section.\arabic{section}.\arabic{subsection}}
  \renewcommand{\theHsubsubsection}{supp.section.\arabic{section}.\arabic{subsection}.\arabic{subsubsection}}
  \renewcommand{\theHfigure}{supp.figure.\arabic{figure}}
  \renewcommand{\theHtable}{supp.table.\arabic{table}}
  \renewcommand{\theHequation}{supp.equation.\arabic{equation}}
}{}
\makeatother

\setcounter{section}{0}
\setcounter{subsection}{0}
\setcounter{subsubsection}{0}
\setcounter{figure}{0}
\setcounter{table}{0}
\setcounter{equation}{0}

\renewcommand{\thesection}{S\arabic{section}}
\renewcommand{\thesubsection}{\thesection.\arabic{subsection}}
\setcounter{section}{0}
\setcounter{figure}{0}
\setcounter{table}{0}
\setcounter{equation}{0}
\renewcommand{\thefigure}{S\arabic{figure}}
\renewcommand{\thetable}{S\arabic{table}}
\renewcommand{\theequation}{S\arabic{equation}}
\setcounter{page}{1}

\begin{center}
  \vspace*{-4pt}
  {\Large \bf Scene-Level Heterogeneous Physics Simulation with 3D Gaussian Splats \par}
  \vspace{10pt}

  {\large  Supplementary Material  \par}
  \vspace{24pt}
\end{center}

\noindent {\Large \textbf{Overview}}
\vspace{0.8em}

\noindent In this supplementary material, we provide additional technical details and results.
Section~\ref{sec:filling} details the particle initialization process;
Section~\ref{sec:consistency} derives the physical consistency of Gaussian covariance recoupling;
Section~\ref{sec:complexity} provides the performance analysis;
Section~\ref{sec:solvers} describes our physics solvers, and Section~\ref{sec:coupling} explains the coupling mechanisms.

\vspace{3em}

\noindent {\Large \textbf{Contents}}
\vspace{1.5em}

\noindent \textbf{\ref{sec:filling}.} \quad Particle initialization \dotfill \\
\noindent \textbf{\ref{sec:consistency}.} \quad Physical consistency of Gaussian covariance recoupling \dotfill \\
\noindent \textbf{\ref{sec:complexity}.} \quad Time complexity and performance analysis \dotfill \\
\noindent \textbf{\ref{sec:solvers}.} \quad Physics simulation solvers \dotfill \\
\noindent \textbf{\ref{sec:coupling}.} \quad The coupling mechanism of unified simulation kernel \dotfill

\vfill

\clearpage
\twocolumn

\section{Particle initialization}
\label{sec:filling}

\subsection{Particle filling for Gaussian splatting}
To enable the physical simulation of 3D Gaussian Splatting (3DGS) assets, which are inherently surface-focused representations, we employ a robust internal particle filling mechanism to generate the necessary volumetric data. Following prior methodologies, we first define the continuous opacity field $d(x)$ derived from the static GS representation $\mathcal{G}_{static}$:
$$
d(x)=\sum_{g\in\mathcal{G}_{static}}\sigma_{g}\exp\left(-\frac{1}{2}(x-\mathbf{k}_{g})^{\mathrm{T}}\mathbf{\Sigma}_{g}^{-1}(x-\mathbf{k}_{g})\right)
$$
This continuous field is discretized onto a dense Eulerian grid with a cell size of $\Delta x$. We identify the interior region of the object using a user-defined opacity threshold $\sigma_{\mathrm{th}}$. Specifically, a grid cell $i$ is considered internal if its center opacity $d(\mathbf{x}_i) > \sigma_{\mathrm{th}}$.

Once the internal volume is identified, we generate the physical particle set $\mathcal{X}_{\mathrm{GS}}$ by randomly sampling positions $\mathbf{x}_j$ within the designated internal grid cells. We employ a Poisson Disk Sampling strategy to ensure a quasi-uniform distribution of particles, preventing clustering and enhancing simulation stability. The minimum sampling distance is set to $r_{\mathrm{min}} = 0.5 \Delta x$.

Crucially, to address potential density disparities during multi-physics coupling, we implement a density homogenization step. If the native particle resolution of the 3DGS asset is significantly lower than that of interacting external solvers (e.g., high-resolution fluids), numerical instability or boundary leakage may occur. In such cases, we inject auxiliary ghost particles into the internal volume to match the target simulation density. These ghost particles participate fully in the physical time integration to enforce robust contact constraints but are explicitly excluded from the rendering pipeline.

\subsection{Physics property allocation}
\label{sec:property_allocation}
Each sampled physical particle $p_j \in \mathcal{X}_{\mathrm{GS}}$, including the ghost particles generated for stability, must be assigned kinematic and constitutive properties for the Material Point Method (MPM) simulation.

\begin{itemize}
    \item \textbf{Initial position and velocity:} The initial position $\mathbf{x}_j^0$ is the sampled position. The initial velocity $\mathbf{v}_j^0$ is set to zero or inherited from a kinematic input.
    \item \textbf{Mass and volume:} We assume the object has a constant, user-specified material density $\rho_{\mathrm{mat}}$. The initial volume $V_j^0$ of each particle is determined by the total volume of the object's occupied space $V_{\mathrm{total}}$ divided by the total number of particles $N_{\mathrm{total}}$: $V_j^0 = V_{\mathrm{total}} / N_{\mathrm{total}}$. The mass is then calculated as $m_j = \rho_{\mathrm{mat}} V_j^0$. $V_{\mathrm{total}}$ is approximated by summing the volume of all internal grid cells identified in Section \ref{sec:filling}.
    \item \textbf{Initial deformation gradient:} The initial elastic deformation gradient is set to the identity matrix: $\mathbf{F}_j^{\mathrm{E},0} = \mathbf{I}$.
    \item \textbf{Visual entity recoupling:} For subsequent visual reconstruction, we store the particle's corresponding rest-state Gaussian covariance $\mathbf{\Sigma}_{\mathrm{rest},j}$ in the Visual Asset Cache. $\mathbf{\Sigma}_{\mathrm{rest},j}$ is either initialized from the nearest existing GS kernel or set as an isotropic sphere $\mathrm{diag}(r_{p}^{2},r_{p}^{2},r_{p}^{2})$ where $r_p$ is the particle radius inferred from $V_j^0$. For ghost particles, this step is skipped to ensure they remain invisible.
\end{itemize}

\section{Physical consistency of Gaussian covariance recoupling}
\label{sec:consistency}

The visual recoupling of the 3D Gaussian covariance matrix $\Sigma$ is derived from the principles of continuum mechanics, specifically how a local affine transformation affects a Gaussian function.

A 3D Gaussian kernel $G(\mathbf{X})$ in the material space $\Omega^0$ is defined by its center $\mathbf{X}_p$ and covariance matrix $A_p$ (denoted as $\Sigma_{rest}$ in our notation):
$$G(\mathbf{X}) = e^{-\frac{1}{2}(\mathbf{X}-\mathbf{X}_p)^T A_p^{-1}(\mathbf{X}-\mathbf{X}_p)}$$
Under a continuous deformation map $\mathbf{x} = \phi(\mathbf{X}, t)$, the deformed kernel $G(\mathbf{x}, t)$ in the world space $\Omega^t$ is:
$$G(\mathbf{x}, t) = e^{-\frac{1}{2}(\phi^{-1}(\mathbf{x}, t)-\mathbf{X}_p)^T A_p^{-1}(\phi^{-1}(\mathbf{x}, t)-\mathbf{X}_p)}$$
Following PhysGaussian, we assume the material point undergoes a local affine transformation characterized by the first-order Taylor expansion around $\mathbf{X}_p$:
$$\tilde{\phi}(\mathbf{X}, t) \approx \mathbf{x}_p + \mathbf{F}_p(\mathbf{X}-\mathbf{X}_p)$$
where $\mathbf{F}_p = \nabla_{\mathbf{X}}\phi(\mathbf{X}, t)$ is the deformation gradient at point $\mathbf{X}_p$.
By inverting this affine map and substituting it back into the exponent, the deformed kernel $\tilde{G}(\mathbf{x}, t)$ remains a Gaussian distribution in the world space $\Omega^t$ with a new center $\mathbf{x}_p$ and a new covariance matrix $\mathbf{a}_p$ (denoted as $\Sigma'$ in our notation):
$$\tilde{G}(\mathbf{x}, t) = e^{-\frac{1}{2}(\mathbf{x}-\mathbf{x}_p)^T (\mathbf{F}_p A_p \mathbf{F}_p^T)^{-1}(\mathbf{x}-\mathbf{x}_p)}$$
This derivation provides the physically grounded rule for updating the covariance matrix:
$$\Sigma' = \mathbf{F} \Sigma_{rest} \mathbf{F}^T$$
This ensures that the shape and orientation of the 3D Gaussian ellipsoids evolve consistently with the underlying continuum mechanics, preserving the non-rigid deformation (stretch and shear) induced by the simulation.

\section{Time complexity and performance analysis}
\label{sec:complexity}

\noindent \textbf{Rendering workflow and comparison.}
A core contribution of our framework is the seamless integration of physically augmented 3D Gaussian Splatting (3DGS) assets into industrial-grade rendering ecosystems, specifically Unreal Engine 5 (UE5). Unlike prior methods that rely on standalone, Python-based rasterizers for visualization, our pipeline involves exporting the deformed states to UE5 to leverage advanced features such as global illumination and complex shadowing. Since this workflow includes manual operations (e.g., asset importation, scene composition, and sequence rendering in UE5), a direct runtime comparison with end-to-end, code-only rendering methods is neither feasible nor meaningful. Our focus is on achieving cinematic-quality offline rendering rather than real-time frame rates.

\noindent \textbf{Simulation costs.}
The computational cost of the physical simulation stage is inherently variable, governed largely by the user-defined simulation precision (grid resolution) and the complexity of the multi-physics coupling (e.g., particle density required for stability). However, the simulation remains efficient for offline content creation. For all five main demonstration scenarios presented in the paper, the physical simulation process was completed in under 10 minutes per sequence on a standard workstation.

\noindent \textbf{Summary.}
We explicitly clarify that our method does not target real-time interactivity. Instead, the primary design goal is to unlock high-fidelity, heterogeneous interactions between captured 3DGS scenes and diverse virtual assets (meshes, particles)—scenarios that require complex solver coupling and are best served by an offline, high-quality rendering workflow.

\section{Physics simulation solvers}
\label{sec:solvers}

\subsection{Rigid body dynamics}
\label{subsec:rigid_solver}
The Rigid Body Solver is designed to simulate articulated mechanisms and multi-body systems using the Reduced Coordinate formulation (also known as Generalized Coordinates). Unlike maximal coordinate methods that treat every link as a free body constrained by algebraic equations, our solver parameterizes the system using joint angles $\mathbf{q} \in \mathbb{R}^{n_{dof}}$ and joint velocities $\dot{\mathbf{q}} \in \mathbb{R}^{n_{dof}}$. This approach naturally satisfies joint constraints and significantly improves computational efficiency for kinematic chains.

\subsubsection{Equations of motion}
The dynamics of the articulated system are governed by the standard manipulator equation:
\begin{equation}
    \mathbf{M}(\mathbf{q}) \ddot{\mathbf{q}} + \mathbf{C}(\mathbf{q}, \dot{\mathbf{q}}) \dot{\mathbf{q}} + \mathbf{g}(\mathbf{q}) = \boldsymbol{\tau}_{ctrl} + \boldsymbol{\tau}_{ext} + \boldsymbol{\tau}_{c}
\end{equation}
where:
\begin{itemize}
    \item $\mathbf{M}(\mathbf{q})$ is the symmetric, positive-definite joint-space inertia matrix (Mass Matrix).
    \item $\mathbf{C}(\mathbf{q}, \dot{\mathbf{q}}) \dot{\mathbf{q}}$ represents the Coriolis and centrifugal forces.
    \item $\mathbf{g}(\mathbf{q})$ is the gravity vector generalized to joint space.
    \item $\boldsymbol{\tau}_{ctrl}$, $\boldsymbol{\tau}_{ext}$, and $\boldsymbol{\tau}_{c}$ are the torques/forces from actuation, external disturbances, and contact constraints, respectively.
\end{itemize}
In our implementation, the bias terms $\mathbf{h}(\mathbf{q}, \dot{\mathbf{q}}) = \mathbf{C}\dot{\mathbf{q}} + \mathbf{g}$ are computed efficiently using recursive kinematics.

\subsubsection{Composite rigid body algorithm (CRBA)}
To compute the mass matrix $\mathbf{M}(\mathbf{q})$, we employ the Composite Rigid Body Algorithm (CRBA). The algorithm operates in two passes. First, it computes the composite inertia $I_i^c$ for each link $i$, which represents the inertia of the subtree rooted at $i$ as if all joints in the subtree were locked:
\begin{equation}
    I_i^c = I_i + \sum_{j \in children(i)} {}^{i}\mathbf{X}_{j}^* I_j^c {}^{j}\mathbf{X}_{i}
\end{equation}
where ${}^{j}\mathbf{X}_{i}$ is the spatial transform from link $i$ to $j$. Second, the entries of $\mathbf{M}$ are populated by projecting the composite inertias onto the joint motion subspaces. This typically achieves $O(N^2)$ complexity but is highly optimized for parallel execution on GPUs.

\subsubsection{Implicit integration and linear solver}
To ensure stability under high stiffness (PD control) and damping, we implement a semi-implicit integration scheme (often referred to as ``Implicit Fast''). We explicitly modify the mass matrix to include the contributions of damping $D$ and stiffness $K$ matrices:
\begin{equation}
    \tilde{\mathbf{M}} = \mathbf{M} + \mathbf{D} \Delta t + \mathbf{K} \Delta t^2
\end{equation}
The discrete-time linear system $\tilde{\mathbf{M}} \Delta \dot{\mathbf{q}} = \tilde{\boldsymbol{\tau}} \Delta t$ is then solved to find the change in velocity. We utilize a dense $LDL^T$ Cholesky decomposition to factorize $\tilde{\mathbf{M}}$:
\begin{equation}
    \tilde{\mathbf{M}} = \mathbf{L} \mathbf{D}_{diag} \mathbf{L}^T
\end{equation}
Forward and backward substitution steps are then performed to solve for accelerations $\ddot{\mathbf{q}}$. This method provides robust behavior for stiff robotic mechanisms without requiring excessively small time steps.

\subsubsection{Kinematics and constraints}
Forward kinematics are computed recursively from the root to leaves, updating the Cartesian position $\mathbf{x}$ and orientation $\mathbf{R}$ of each link. Constraint forces $\boldsymbol{\tau}_{c}$ arising from collisions and joint limits are solved using an impulse-based approach or Projected Gauss-Seidel (PGS) solver, integrated tightly with the forward dynamics loop.

\subsection{Material point method (MPM)}
\label{subsec:mpm_solver}
The Material Point Method (MPM) solver in our unified kernel serves as the primary engine for simulating continuum materials, including elastoplastic solids, granular media, and fluids. We implement the Moving Least Squares MPM (MLS-MPM) formulation combined with the Affine Particle-in-Cell (APIC) transfer scheme to ensure energy conservation and reduce numerical dissipation.

\subsubsection{Discretization and state variables}
The continuum body is discretized into a set of Lagrangian particles $p$, each carrying mass $m_p$, position $\mathbf{x}_p$, velocity $\mathbf{v}_p$, deformation gradient $\mathbf{F}_p$, and an affine velocity field matrix $\mathbf{C}_p$. A background Eulerian Cartesian grid with cell size $\Delta x$ is employed as a scratchpad for computing spatial gradients and solving the equations of motion.

\subsubsection{Particle-to-grid transfer (P2G)}
At the beginning of each time substep $\Delta t$, mass and momentum are transferred from particles to grid nodes $i$ using quadratic B-spline shape functions $N_i(\mathbf{x}_p)$. Under the APIC formulation, the momentum transfer includes the affine velocity contribution to preserve angular momentum:
\begin{equation}
    m_i = \sum_p N_i(\mathbf{x}_p) m_p
\end{equation}
\begin{equation}
    (m\mathbf{v})_i = \sum_p N_i(\mathbf{x}_p) m_p \left[ \mathbf{v}_p + \mathbf{C}_p (\mathbf{x}_i - \mathbf{x}_p) \right]
\end{equation}
In our MLS-MPM implementation, the internal forces derived from the material constitutive model are applied directly during this transfer via the affine term. Let $\Psi(\mathbf{F})$ be the elastic energy density. The first Piola-Kirchhoff stress $\mathbf{P}(\mathbf{F}) = \partial \Psi / \partial \mathbf{F}$ is computed and fused into the momentum update to avoid explicit grid force accumulation:
\begin{equation}
    (\mathbf{v}_i)^{new} = \frac{(m\mathbf{v})_i}{m_i} + \Delta t \mathbf{g} - \frac{\Delta t}{m_i} \sum_p V_p \mathbf{P}(\mathbf{F}_p)^T \nabla N_i(\mathbf{x}_p)
\end{equation}
where $V_p$ is the particle volume and $\mathbf{g}$ is the gravity vector.

\subsubsection{Constitutive model and plasticity}
To handle finite deformations and plasticity, we adopt a Singular Value Decomposition (SVD) based return-mapping algorithm. The trial deformation gradient is updated via the velocity gradient:
\begin{equation}
    \mathbf{F}_p^{trial} = (\mathbf{I} + \Delta t \mathbf{C}_p) \mathbf{F}_p^n
\end{equation}
We decompose $\mathbf{F}_p^{trial} = \mathbf{U}_p \boldsymbol{\Sigma}_p \mathbf{V}_p^T$. Plasticity is applied by clamping the singular values $\boldsymbol{\Sigma}_p$ based on the specific material model (e.g., Drucker-Prager for sand, Von Mises for metal, or Fixed Corotated for elastic tissues), yielding the projected deformation gradient $\mathbf{F}_p^{n+1}$.

\subsubsection{Grid-to-particle transfer (G2P)}
After solving the grid velocities, the updated states are transferred back to the particles. The particle velocity and the affine velocity field are updated as follows:
\begin{equation}
    \mathbf{v}_p^{n+1} = \sum_i N_i(\mathbf{x}_p) \mathbf{v}_i^{new}
\end{equation}
\begin{equation}
    \mathbf{C}_p^{n+1} = \frac{4}{\Delta x^2} \sum_i N_i(\mathbf{x}_p) \mathbf{v}_i^{new} (\mathbf{x}_i - \mathbf{x}_p)^T
\end{equation}
Finally, particle positions are advected: $\mathbf{x}_p^{n+1} = \mathbf{x}_p^n + \Delta t \mathbf{v}_p^{n+1}$.

\subsection{Position-based dynamics (PBD)}
\label{subsec:pbd_solver}
The Position-Based Dynamics (PBD) solver is employed to simulate deformable objects such as cloth, soft bodies, and Lagrangian fluids. Our implementation adheres to the Extended Position-Based Dynamics (XPBD) formulation, which introduces a compliance parameter to decouple material stiffness from time step size and iteration counts.

\subsubsection{Simulation loop}
The solver follows a predictor-corrector scheme. For each particle $i$, a tentative position $\mathbf{x}_i^*$ is first predicted using explicit Euler integration based on external forces $\mathbf{f}_{ext}$ (e.g., gravity):
\begin{equation}
    \mathbf{v}_i^* = \mathbf{v}_i^n + \Delta t \mathbf{M}^{-1} \mathbf{f}_{ext}, \quad \mathbf{x}_i^* = \mathbf{x}_i^n + \Delta t \mathbf{v}_i^*
\end{equation}
Subsequently, the solver iteratively resolves a set of geometric constraints $C(\mathbf{x}) = 0$. The position correction $\Delta \mathbf{x}$ for a constraint involving a set of particles is derived by minimizing the constraint potential energy:
\begin{equation}
    \Delta \lambda = \frac{-C(\mathbf{x}^*) - \tilde{\alpha}\lambda}{\sum_j w_j |\nabla_j C(\mathbf{x}^*)|^2 + \tilde{\alpha}}
\end{equation}
\begin{equation}
    \Delta \mathbf{x}_i = w_i \Delta \lambda \nabla_i C(\mathbf{x}^*)
\end{equation}
where $w_i = 1/m_i$ is the inverse mass, and $\tilde{\alpha} = \alpha / \Delta t^2$ is the discrete compliance derived from the physical compliance $\alpha$. Finally, velocities are updated: $\mathbf{v}_i^{n+1} = (\mathbf{x}_i^{n+1} - \mathbf{x}_i^n) / \Delta t$.

\subsubsection{Structural constraints}
We implement specific constraints for different material types:
\begin{itemize}
    \item \textbf{Distance constraint (cloth/links):} To model stretching resistance, we enforce the distance between two particles $i$ and $j$:
    \begin{equation}
        C_{dist}(\mathbf{x}_i, \mathbf{x}_j) = \|\mathbf{x}_i - \mathbf{x}_j\| - l_0 = 0
    \end{equation}

    \item \textbf{Bending constraint (cloth):} To simulate bending resistance in triangular meshes, we utilize the isometric bending constraint based on the dihedral angle $\phi$ between adjacent triangle faces:
    \begin{equation}
        C_{bend}(\mathbf{x}_1, \mathbf{x}_2, \mathbf{x}_3, \mathbf{x}_4) = \text{acos}(\mathbf{n}_1 \cdot \mathbf{n}_2) - \pi = 0
    \end{equation}

    \item \textbf{Volume constraint (soft bodies):} For tetrahedral meshes, volume conservation is enforced on each element to model material compressibility:
    \begin{equation}
        C_{vol}(\mathbf{x}_1, \mathbf{x}_2, \mathbf{x}_3, \mathbf{x}_4) = \frac{1}{6} (\mathbf{x}_{21} \times \mathbf{x}_{31}) \cdot \mathbf{x}_{41} - V_0 = 0
    \end{equation}
\end{itemize}

\subsubsection{Position-based fluids (PBF)}
For liquid simulation, we adopt the PBF framework which enforces constant density. The density $\rho_i$ at particle $i$ is estimated using the Poly6 kernel $W_{poly6}$:
\begin{equation}
    \rho_i = \sum_j m_j W_{poly6}(\|\mathbf{x}_i - \mathbf{x}_j\|, h)
\end{equation}
The density constraint requires $C_{density}(\mathbf{x}_1, \dots, \mathbf{x}_n) = \frac{\rho_i}{\rho_0} - 1 = 0$. Gradients are computed using the Spiky kernel $\nabla W_{spiky}$ to avoid clustering instability.

Viscosity is handled via XSPH (Artificial viscosity), which smoothes the velocity field before the position update:
\begin{equation}
    \mathbf{v}_i^{new} = \mathbf{v}_i + c \sum_j \frac{m_j}{\rho_j} (\mathbf{v}_j - \mathbf{v}_i) W_{poly6}(\|\mathbf{x}_i - \mathbf{x}_j\|, h)
\end{equation}
where $c$ is a tunable viscosity coefficient.

\subsection{Smoothed particle hydrodynamics (SPH)}
\label{subsec:sph_solver}
The SPH solver provides a fully Lagrangian approach for simulating fluid dynamics, capable of handling free-surface flows and complex topological changes. Our implementation supports both Weakly Compressible SPH (WCSPH) for efficiency in dynamic scenarios and Divergence-Free SPH (DFSPH) for enforcing strict incompressibility with larger time steps.

\subsubsection{Discretization}
Fluid quantities at a particle $i$ are interpolated from its neighbors $j$ using a smoothing kernel $W_{ij} = W(\|\mathbf{x}_i - \mathbf{x}_j\|, h)$ with support radius $h$. The density $\rho_i$ is evaluated via standard summation:
\begin{equation}
    \rho_i = \sum_j m_j W_{ij}
\end{equation}
The momentum equation governing the fluid motion is given by:
\begin{equation}
    \frac{d\mathbf{v}_i}{dt} = \frac{\mathbf{F}_i^{pressure}}{m_i} + \frac{\mathbf{F}_i^{viscosity}}{m_i} + \frac{\mathbf{F}_i^{surface}}{m_i} + \mathbf{g}
\end{equation}

\subsubsection{Pressure solvers}
We provide two distinct formulations to resolve the pressure gradient term $\mathbf{F}^{pressure}$:

\paragraph{Weakly compressible SPH (WCSPH)}
In this formulation, pressure is explicitly computed from density deviations using the Tait equation of state, allowing for slight compressibility (typically $< 1\%$):
\begin{equation}
    p_i = \frac{k \rho_0}{\gamma} \left( \left( \frac{\rho_i}{\rho_0} \right)^\gamma - 1 \right)
\end{equation}
where $k$ is the stiffness constant and $\gamma=7$. The resulting pressure force is symmetric to ensure momentum conservation:
\begin{equation}
    \mathbf{F}_i^{pressure} = -m_i \sum_j m_j \left( \frac{p_i}{\rho_i^2} + \frac{p_j}{\rho_j^2} \right) \nabla W_{ij}
\end{equation}

\paragraph{Divergence-free SPH (DFSPH)}
For scenarios requiring stiff incompressibility, we implement DFSPH, which enforces two constraints using iterative pressure Poisson solvers.
First, a \textit{divergence-free solver} modifies the velocities to ensure $\nabla \cdot \mathbf{v} = 0$ before position integration:
\begin{equation}
    \frac{D\rho}{Dt} = -\rho (\nabla \cdot \mathbf{v}) = 0 \implies \sum_j m_j (\mathbf{v}_i - \mathbf{v}_j) \cdot \nabla W_{ij} = 0
\end{equation}
Second, a \textit{constant-density solver} corrects position updates to maintain $\rho = \rho_0$ (or equivalently, corrects predicted density error). This predictor-corrector scheme significantly improves stability under large time steps compared to WCSPH.

\subsubsection{Viscosity and Surface Tension}
To simulate viscous fluids, we apply a Laplacian-based viscosity force that dampens relative velocities between neighboring particles:
\begin{equation}
    \mathbf{F}_i^{viscosity} = \sum_j \frac{m_j}{\rho_j} \mu \frac{(\mathbf{v}_j - \mathbf{v}_i) \cdot \mathbf{x}_{ij}}{\|\mathbf{x}_{ij}\|^2 + \epsilon h^2} \nabla W_{ij}
\end{equation}
where $\mu$ is the dynamic viscosity coefficient. Additionally, surface tension is modeled as a pairwise cohesive force proportional to a tension coefficient $\gamma_{tension}$:
\begin{equation}
    \mathbf{F}_i^{surface} = -\gamma_{tension} \sum_j m_j (\mathbf{x}_i - \mathbf{x}_j) W(\|\mathbf{x}_i - \mathbf{x}_j\|)
\end{equation}
This term mimics molecular attraction, promoting the formation of droplets and minimizing surface area.

\section{The coupling mechanism of unified simulation kernel}
\label{sec:coupling}
The Unified Simulation Kernel manages heterogeneous interactions by abstracting all assets into particles or meshes and utilizing specialized, coupled solvers. To ensure physical plausibility and conservation laws across different integration schemes, the coupling logic is categorized into three distinct mechanisms based on the solver types involved: (1) Signed Distance Field (SDF) based impulse response for Rigid-MPM/SPH interactions, (2) Position-based penetration correction for Rigid-PBD interactions, and (3) Grid-mediated momentum exchange for MPM-SPH/PBD interactions.

\subsection{SDF-based impulse response (rigid-MPM and rigid-SPH)}
For interactions between rigid bodies and Eulerian-Lagrangian hybrid solvers (MPM) or purely Lagrangian solvers (SPH), a velocity-level impulse method is employed. This approach utilizes the high-resolution SDF of rigid geometries to modulate collision softness and friction.

For a particle $p$ (either an MPM grid node or an SPH particle) located at $\mathbf{x}_p$ with velocity $\mathbf{v}_p$, we first compute the signed distance $d(\mathbf{x}_p)$ and the surface normal $\mathbf{n}$ from the rigid body's geometry. A blending weight $w$, determined by a softness parameter $\epsilon$, allows for smooth contact handling:
\begin{equation}
    w(d) = \min\left(\exp\left(-\frac{d}{\max(\eta, \epsilon)}\right), 1\right)
\end{equation}
where $\eta$ is a small numerical stabilizer.

The relative velocity with respect to the rigid body velocity $\mathbf{v}_r$ is defined as $\mathbf{v}_{rel} = \mathbf{v}_p - \mathbf{v}_r$. We decompose $\mathbf{v}_{rel}$ into normal component $v_n = \mathbf{v}_{rel} \cdot \mathbf{n}$ and tangential vector $\mathbf{v}_t = \mathbf{v}_{rel} - v_n \mathbf{n}$.

If $v_n < 0$ (penetration velocity), the post-collision relative velocity is computed by applying restitution $e$ and Coulomb friction $\mu$. The normal impulse response is $\tilde{v}_n = -e v_n$. The tangential response uses a friction cone constraint:
\begin{equation}
    \tilde{\mathbf{v}}_t = \frac{\mathbf{v}_t}{\|\mathbf{v}_t\|} \max\left(0, \|\mathbf{v}_t\| - \mu |v_n|\right)
\end{equation}
The final updated velocity of the particle is a blend of the rigid body velocity and the new relative velocity, weighted by $w$:
\begin{equation}
    \mathbf{v}_{p}^{new} = \mathbf{v}_r + w(\tilde{\mathbf{v}}_t + \tilde{v}_n \mathbf{n}) + (1-w)\mathbf{v}_{rel}
\end{equation}
Momentum conservation is strictly enforced by applying the reverse impulse $-\Delta \mathbf{P} / \Delta t$ as an external force to the rigid body solver.

\subsection{Position-based penetration correction (rigid-PBD)}
Interactions between rigid bodies and Position-Based Dynamics (PBD) particles rely on geometric projection rather than velocity impulses. This method prioritizes resolving interpenetration immediately at the position level.

For a PBD particle with mass $m_p$ and radius $r$, collision is detected when the signed distance $d(\mathbf{x}_p) < r$. The particle position is projected outward along the contact normal $\mathbf{n}$ to resolve the overlap:
\begin{equation}
    \mathbf{x}_{p}^{new} = \mathbf{x}_{p}^{old} + k (r - d(\mathbf{x}_p)) \mathbf{n}
\end{equation}
where $k \in [0, 1]$ represents the stiffness coefficient. Unlike the impulse-based method, friction is explicitly neglected in this formulation to maintain stability in the position projection step.

The velocity is implicitly updated via the positional change: $\mathbf{v}_{p}^{new} = (\mathbf{x}_{p}^{new} - \mathbf{x}_{p}^{old}) / \Delta t$. The resulting momentum change $\Delta \mathbf{P} = m_p (\mathbf{v}_{p}^{new} - \mathbf{v}_{p}^{old})$ is accumulated and applied as a reaction force on the coupled rigid body to satisfy Newton's third law.

\subsection{Grid-mediated momentum exchange (MPM-SPH and MPM-PBD)}
Coupling between the Material Point Method (MPM) and Lagrangian particles (SPH or PBD) is achieved via the background MPM Eulerian grid, which acts as a momentum exchange medium. This mechanism models a perfectly inelastic collision process within each time substep.

For a given MPM grid node $g$, we identify the set of neighboring external particles $\mathcal{N}_g$ (SPH or PBD) within the local stencil. The algorithm forces a velocity synchronization where the grid velocity $\mathbf{v}_g$ is overridden by the average velocity of the interacting particles:
\begin{equation}
    \mathbf{v}_g^{sync} = \frac{1}{|\mathcal{N}_g|} \sum_{i \in \mathcal{N}_g} \mathbf{v}_{p_i}
\end{equation}
This operation implies an instantaneous ``sticking'' condition. To conserve total momentum, the change in the grid's momentum $\Delta \mathbf{P}_g = m_g (\mathbf{v}_g^{sync} - \mathbf{v}_g^{old})$ is calculated. This momentum difference is then distributed back to the external particles as a corrective feedback:
\begin{equation}
    \mathbf{v}_{p_i}^{new} = \mathbf{v}_{p_i} - \frac{\Delta \mathbf{P}_g}{m_{p_i}}
\end{equation}
This two-way transfer ensures that the MPM fluid effectively drags or is dragged by the immersed SPH/PBD particles, behaving similarly to a mixture model with high drag coefficients.